\documentclass[twocolumn,secnumarabic,amssymb, nobibnotes, aps, prl]{revtex4-1}
\usepackage{graphicx}
\usepackage{textcomp}
\usepackage{amsmath}
\usepackage{commath}
\usepackage{color}
\begin{document}
	
	\title{Nonlinear harmonic generation in two-dimensional lattices of repulsive magnets}
	\author{Weijian Jiao and Stefano Gonella}
	\affiliation{Department of Civil, Environmental, and Geo- Engineering\\University of Minnesota, Minneapolis, MN 55455, USA\\}
	
	\begin{abstract}
	In this Letter, we provide experimental evidence of nonlinear wave propagation in a triangular lattice of repulsive magnets supported by an elastic foundation of thin pillars and we interpret all the individual features of the nonlinear wavefield through the lens of a phonon band calculation that precisely accounts for the inter-particle repulsive forces. We confirm the co-existence of two spectrally distinct components (homogeneous and forced) in the wave response that is induced via second harmonic generation (SHG), a well-known effect of quadratic nonlinearity (here embedded in the magnetic interaction). We show that the modal and spatial characteristics of the second harmonic components are complementary to those exhibited by the fundamental harmonic. This endows the lattice with a functionality enrichment capability, whereby additional modes and directivity patterns can be triggered and tuned by merely increasing the amplitude of excitation.
		
		\vspace{0.4cm}
	\end{abstract}

	\maketitle
    
    In recent years, nonlinear periodic structures and acoustic metamaterials have been extensively studied because of their rich dynamical behavior and for their tunability and adaptivity characteristics. A number of studies have focused on the propagation of solitary waves and discrete breathers in a variety of material systems, such as granular crystals \cite{Daraio_2005, Daraio_2006, Leonard2013, Boechler_2010}, magnetic systems \cite{Russell_1997, Miguel_2014, Moler_n_2019}, and mechanical metamaterials \cite{Chen_2014, Deng_2017_PRL, Deng_2018}. Other notable works have explored metastructures equipped with bistable or bucklable elements and exhibiting tuning and energy harvesting functionalities \cite{Mullin_2007, Wang_2014, Neel_2016, Shan_2015}. In general, achieving these nonlinear effects requires activating the strongly nonlinear response associated with large deformation that is achievable, for instance, working with soft materials or thin structures. 
    
    Tuning effects can also be triggered and harnessed in weakly nonlinear systems, and their implications have been investigated for wave control of phononic media. In the case of weak cubic nonlinearity, the main manifestation of nonlinearity is an amplitude-dependent correction of the dispersion relation, which, in principle, enables shifting the onset and width of bandgaps via a simple control of the excitation amplitude \cite{Narisetti_2010, Narisetti_2011, Cabaret_2012_PRE, Bonanomi_2015}. Recently, this tuning effect has been employed to control edge states in topological phononic lattices \cite{Pal_2018}. Another weakly nonlinear effect of great relevance is the second harmonic generation (SHG), which is the main signature of quadratic nonlinearity - a dominant contribution in many nonlinear physical systems \cite{tournat2004, Matlack_2011, Cabaret_2012_PRE, Tournat_2013, Mehrem_2017, JIAO2018, Jiao_2018_PRA, Ganesh_2017, wallen2017, GRINBERG_2020}. Harnessing SHG in nonlinear acoustic metamaterials has opened new doors for a broad range of applications, including acoustic diodes and switches \cite{Liang_2010, Boechler_2011}, subwavelength energy trapping \cite{Jiao_2018_PRA}, and adaptive spatial directivity \cite{Ganesh_APL_2017}. The opportunity spectrum gets even wider if we consider systems that feature simultaneously cubic and quadratic nonlinearities, where the correction of the band diagram induced by cubic nonlinearity affects indirectly the manifestation of SHG, providing a secondary tuning capability, as recently shown in \cite{Jiao_2019_PRE}.
    
    While SHG has been widely studied in one-dimensional nonlinear metamaterials and waveguides, the investigation of its effects on the spatial characteristics of nonlinear wavefields in 2D metamaterials has been more sporadic \cite{Ganesh_2017, Ganesh_APL_2017} and still lacks a definitive experimental observation of nonlinear response at amplitude levels that are suitable for practical applications. In this Letter, we attempt to bridge this gap by experimentally investigating SHG in a discrete system consisting of a periodic network of repulsively interacting magnets supported by an elastic foundation of thin pillars. The system can be interpreted as a practical realization of a triangular lattice of particles with on-site potentials. Specifically, we first verify the existence of SHG in the spectrum of the response. To this end, we take advantage of the strength of the nonlinear effects granted by the particle nature of the system, compared to the case of structural lattices, even without the establishment of phase matching conditions. Moreover, we experimentally confirm that the second harmonic encompasses two contributions, traditionally referred to as the forced and homogeneous components \cite{Ganesh_2017, JIAO2018}. Finally, we show that the second harmonic features distinctive and complementary modal and spatial characteristics when we compare the response with that of the fundamental harmonic. 
    
    \begin{figure} [!htb]
	\centering
	\includegraphics[scale=0.06]{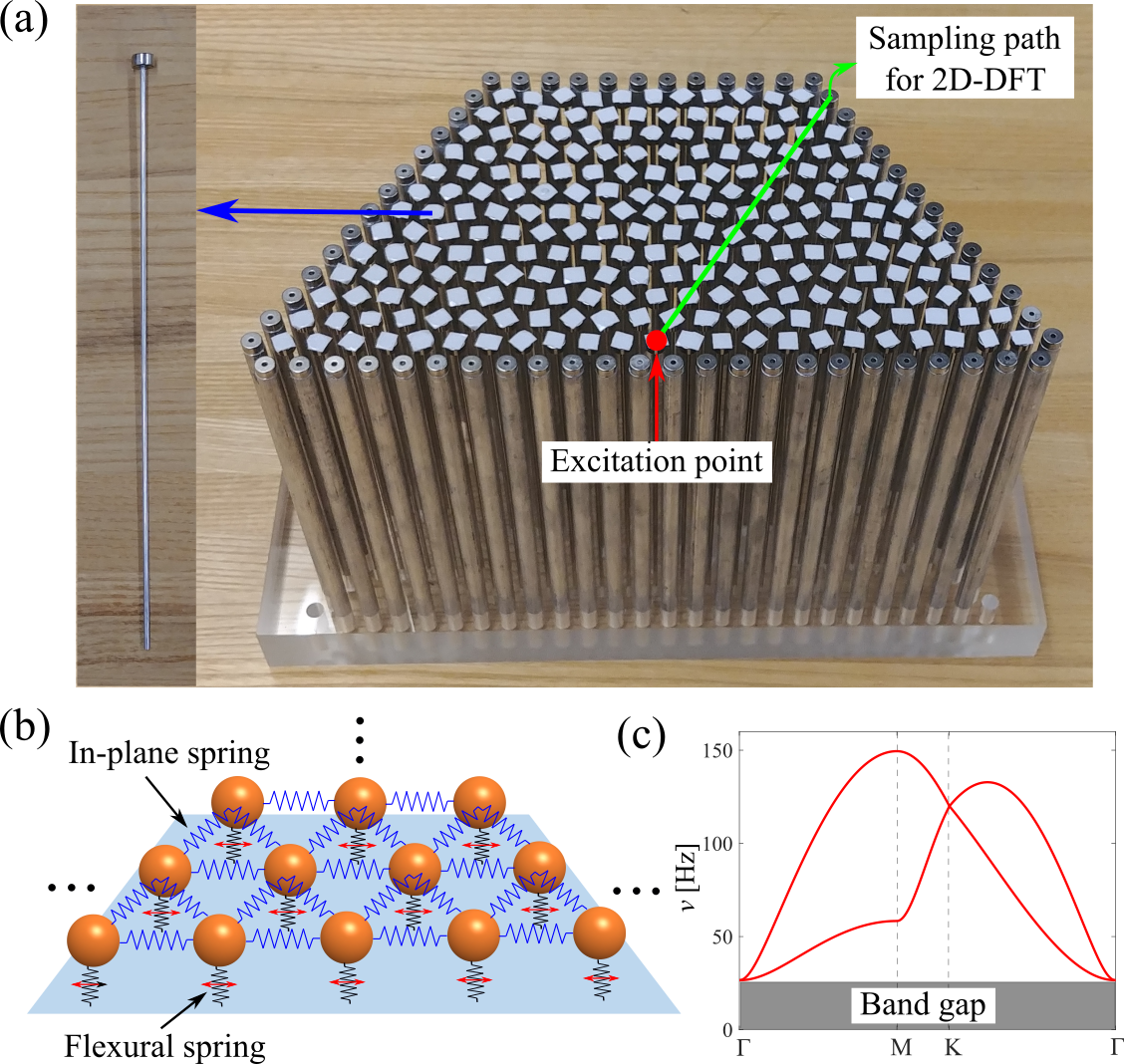}
	\caption{(a) Lattice specimen consisting of magnets supported by thin beams in the interior (shown in the inset) and thick beams along the boundary. The interior magnets are covered by reflective tape to enhance laser measurements. (b) Equivalent spring-mass model of the lattice, showing the function of the pillars acting as an elastic foundation. (c) Linear dispersion relation obtained from the analytical model in \cite{jiao2019_arXiv}, accounting for the effect of the static inter-particle repulsive forces.}
	\label{Specimen_Model}
    \end{figure}

    The specimen used for these tasks, shown in Fig.~\ref{Specimen_Model}(a), consists of an array of pillars arranged to form a triangular lattice occupying a half-hexagon domain. Each pillar consists of a magnetic ring (Grade N42, with 1/4 inch outer diameter, 1/16 inch inter diameter and 1/8 inch thickness) inserted at the tip of a slender Aluminum beam whose other end is plugged into an Acrylic base through a drilled hole. The pillars in the interior of the lattice feature slender beams (1/16 inch in diameter).  For the exterior pillars, the magnets are simply glued to the tips of thick beams (1/4 inch in diameter) featuring large bending stiffness to effectively establish fixed boundary conditions along the contour of the hexagonal domain. The magnets are arranged as to experience side-by-side repulsive interaction in their own plane, and each magnet is initially in equilibrium under the action of the self-balancing static forces exerted by its neighbors. The configuration guarantees that, for the amplitudes of interest for this study, the motion of the magnets remains confined within the plane of the lattice. This setup is imported and adapted from a previous experimental effort, in which we used this platform to characterize the linear response of lattices of magnetically interacting particle systems \cite{jiao2019_arXiv}. Through those experiments, we were able to demonstrated a series of non-intuitive correction effects induced on the lattice band structure by the static inter-particle repulsive forces. In this work, we leverage these key results as a precious guideline for the nonlinear investigation. 
    	
    \begin{figure*} [!htb]
    	\centering
    	\includegraphics[scale=0.6]{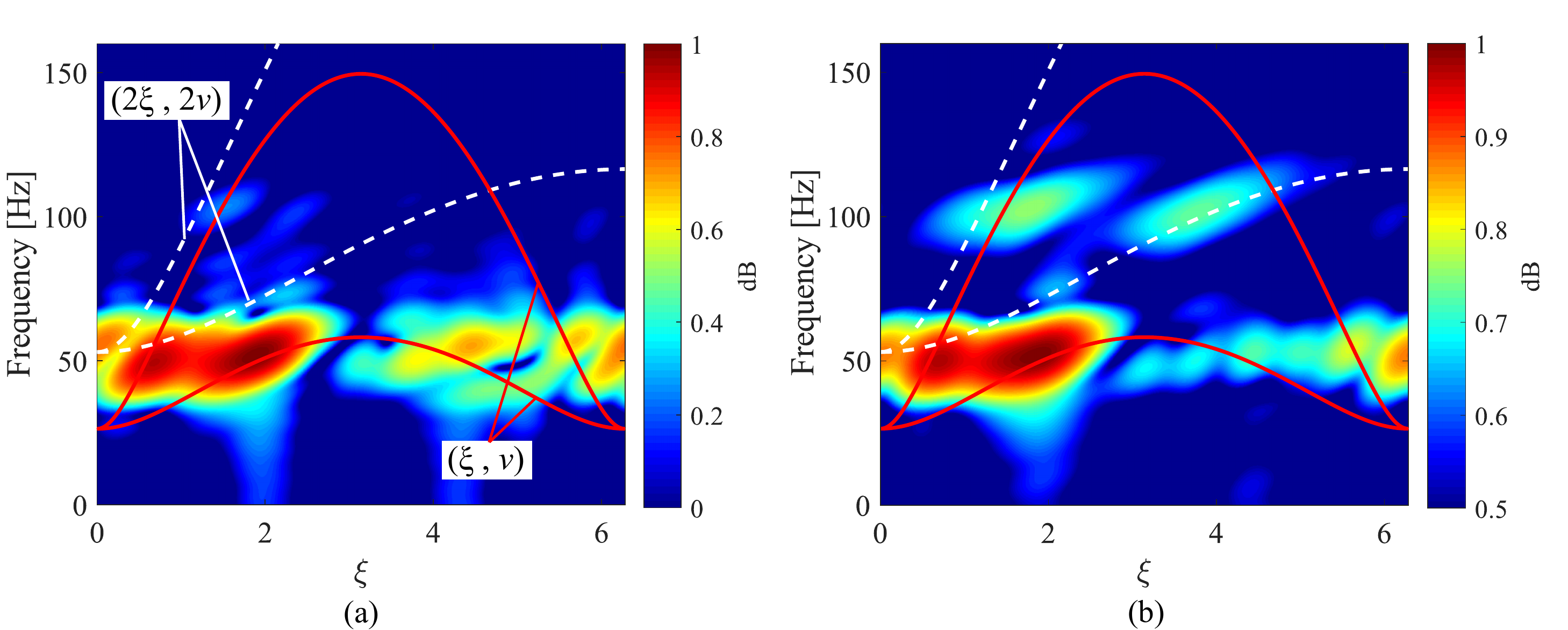}
    	\caption{Spectra of the experimental responses for tone-burst excitations at 50 Hz. (a) Response to low-amplitude excitation, showing the fundamental harmonic activating both shear and longitudinal modes (red solid curves). (b) Response to high-amplitude excitation, showing unequivocally two distinct signatures at the second harmonic (i.e., 100 Hz), confirming the existence of both homogeneous and forced components of the nonlinearly generated harmonic. The white dashed curves represent all the possible spectral points that can be activated by the forced component.}
    	\label{Spectrum_50Hz_nonlinear}
    \end{figure*}

    As shown in Fig.~\ref{Specimen_Model}(b), the system is modeled as a triangular spring-mass lattice in which each node is connected to ground through a flexural spring that captures the elastic foundation effect of the supporting beam. The in-plane repulsive interaction between neighboring magnets is modeled as a nonlinear spring featuring an inverse power law $f(r)=\beta r^{-\alpha}$, with $\alpha = 4.5824$ and $\beta = 1.6209 \times 10^{-10}$. Here, the parameters have been obtained by fitting the  force-displacement relation between two magnets acquired experimentally through a micrometer equipped with a highly sensitive load cell (see details in the SI of \cite{jiao2019_arXiv}). The spring constant of the flexural springs in the foundation is taken as the equivalent flexural stiffness of a cantilever beam with the cross sectional and material properties of the pillar, and found to be $k_f=19.6757$ N/m. As for the other parameters, $m=7.07 \times 10^{-4}$ kg is the mass of each magnet and $L_0=0.01$ m is the initial spacing between two nodes in the lattice. In our previous study \cite{jiao2019_arXiv}, we have shown that the correct dispersive properties of repulsive lattices cannot be resolved using conventional harmonic spring-mass models. Specifically, we have demonstrated that the static repulsive forces due to the magnets, which constraint the lattice at equilibrium, can induce large softening corrections of dispersion branches - an effect especially felt by the shear mode. Adapting the result in \cite{jiao2019_arXiv}, the band diagram of a repulsive lattice with an elastic foundation can be obtained by solving the following eigenvalue problem
    \begin{equation}\label{Eigenvalue_prob}
    \left[ -\omega^2\mathbf{M} +\mathbf{D}(\mathbf{k})\right] \boldsymbol{\phi}=\mathbf{0}
    \end{equation}
    where $\omega=2\pi \nu$, $\mathbf{k}$ is the wavevector, $\mathbf{M}=\begin{bmatrix} m & 0 \\ 0 & m \end{bmatrix}$ is the mass matrix and \begin{multline}\label{Stiffness_matrix}
    \mathbf{D}(\mathbf{k})=2\sum_{l=1}^{3}\left\lbrace  f_r(L_0)\mathbf{e}_l\otimes    \mathbf{e}_l\left[ \cos(\mathbf{k} \cdot \mathbf{R}_l)-1\right]  \right\rbrace \\ + 2\sum_{l=1}^{3}\left\lbrace\frac{f(L_0)}{L_0}\left( \mathbf{I}-\mathbf{e}_l \otimes \mathbf{e}_l\right) 
    \left[ \cos(\mathbf{k} \cdot \mathbf{R}_l)-1\right]  \right\rbrace + \mathbf{K}_f
    \end{multline}
    is a wavevector-dependent dynamical matrix that already incorporates Bloch conditions, and $\mathbf{R}_1=L_0\mathbf{e}_1=L_0[1 \,\ 0 ]^T$, $\mathbf{R}_2=L_0\mathbf{e}_2=L_0[1/2 \,\ \sqrt{3}/2]^T$, and $\mathbf{R}_3=L_0\mathbf{e}_3=L_0[-1/2 \,\ \sqrt{3}/2]^T$ are the lattice vectors. In Eq.~\ref{Stiffness_matrix}, the first term is the conventional stiffness matrix for harmonic particle systems, where $f_r(L_0)$ is the first derivative of the repulsive force $f(r)$ evaluated at the initial nodal spacing $L_0$, the second term is an additional stiffness contribution capturing the aforementioned softening effect and $\mathbf{K}_f=\begin{bmatrix} k_{f} & 0 \\ 0 & k_{f} \end{bmatrix}$ represents the stiffness of the elastic foundation. The solution of Eq.~\ref{Eigenvalue_prob} yields the band diagram plotted in Fig.~\ref{Specimen_Model}(c). As expected, the band diagram is fully gapped at low frequencies as a result of the elastic foundation, and we observe the two canonical acoustic branches, the first is dominated by shear mechanisms and the second by longitudinal mechanisms. 

	\begin{figure*} [!htb]
	\centering
	\includegraphics[scale=0.16]{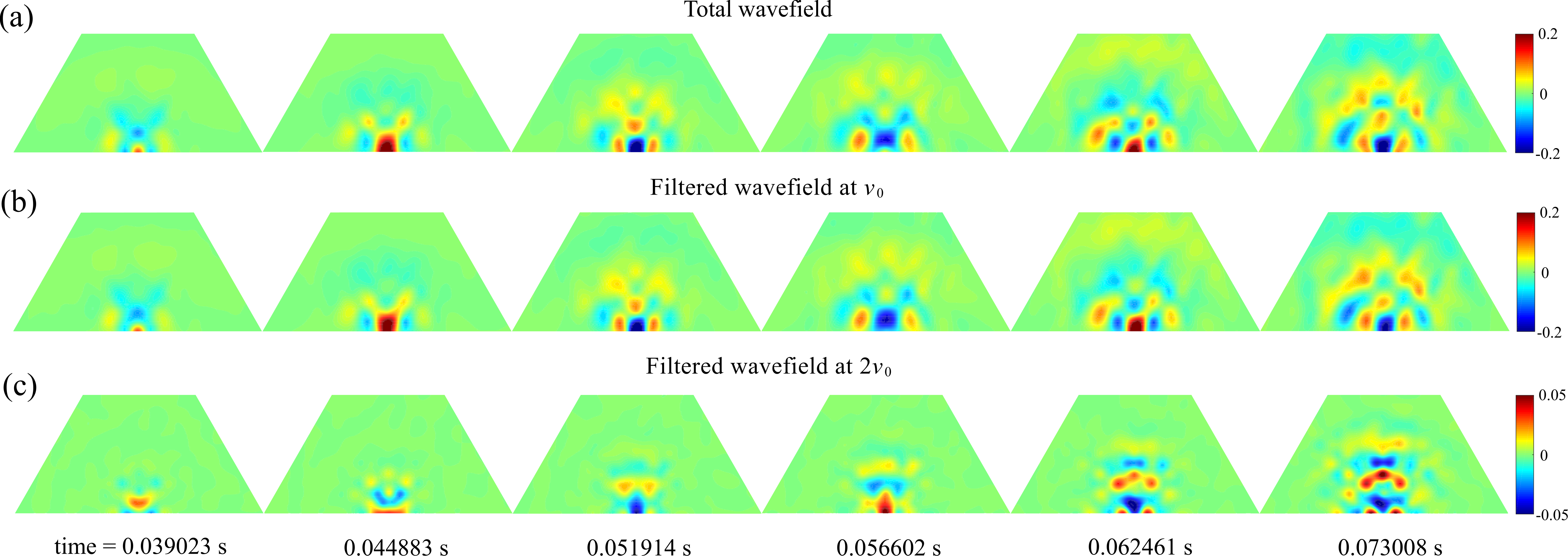}
	\caption{Snapshots of the wavefields experimentally acquired by laser scans at six successive time instants, showing distinct spatial patterns at the different harmonics. (a) Total wavefield. (b) Wavefield filtered at $\nu_0$, highlighting the fundamental harmonic. (c) Wavefield filtered at $2\nu_0$, highlighting the second harmonic. The two harmonics feature complementary modal and directional characteristics, thus exposing the functionality enrichment that is achieved by triggering a nonlinear response.}
	\label{Wavefield}
    \end{figure*}

	We now proceed to investigate the nonlinear response. To experimentally capture the in-plane response of the lattice, we employ a 3D scanning laser Doppler vibrometer (SLDV, Polytec PSV-400-3D) by which we measure the displacement of the individual magnets. The specimen is excited in the vertical direction by a force applied to the center magnet of the bottom layer (as indicated by the red dot in Fig.~\ref{Spectrum_50Hz_nonlinear}(a)) through a Bruel \& Kjaer Type $4809$ shaker, powered by a Bruel \& Kjaer Type $2718$ amplifier. The excitation is prescribed as a five-cycle tone burst with carrier frequency $\nu_0=50$ Hz. First, we use a low-amplitude excitation to elicit a linear response of the specimen. In Fig.~\ref{Spectrum_50Hz_nonlinear}(a) we plot the color map of the normalized spectral amplitude obtained via 2D discrete Fourier transform (2D-DFT) in space and time of the experimental spatio-temporal response sampled along one lattice vector (i.e., along the green line shown in Fig.~\ref{Specimen_Model}(a)). Then, we progressively raise the amplitude of excitation until the deformations are sufficiently large to activate non-negligible nonlinearity in the response, and the corresponding spectral amplitude map is given in Fig.~\ref{Spectrum_50Hz_nonlinear}(b). In both figures, we superimpose the $\mathrm{\Gamma}$-$\mathrm{M}$ portion of the linear dispersion branches (red curves) obtained via Bloch analysis informed using our modified lattice model and periodically extended into the second Brillouin zone for convenience. In addition, the white dashed curves denote the parametric locus of the $2\xi$ - $2\nu_{1,2}(\xi)$ pairs, i.e., the spectral points that feature simultaneously twice the frequency and twice the wavenumber of the acoustic phonons at the fundamental harmonic $(\xi,\nu_{1,2}(\xi))$. The $2\xi$ - $2\nu_{1,2}(\xi)$ points represent the pairs of frequency and wavenumber that can be displayed by the forced component of a nonlinearly generated second harmonic. Since these pairs do not live on any dispersion branch (and therefore do not conform to any mode of the linear system), their activation is conditional upon the generation of harmonics and is therefore the most robust detector of nonlinearity in the wave response. In other words, phonons that live on these curves must be generated through nonlinear mechanisms intrinsic to the lattice and cannot be merely induced by an external excitation prescribed at $2\nu$. This consideration provides a powerful tool to distinguish with absolute certainty the manifestation of nonlinearly generated harmonics from other spurious signatures of high-frequency components that could be already embedded in the excitation signal (for example due to nonlinearities in the signal generation and amplification). As expected, in Fig.~\ref{Spectrum_50Hz_nonlinear}(a) we observe that the main spectral contribution is located at the prescribed frequency ($\nu_0= 50$ Hz), and both shear and longitudinal modes are activated, with no appreciable signature at the second harmonic ($2\nu_0=100$ Hz). In contrast, in the spectrum of the nonlinear response (i.e., Fig.~\ref{Spectrum_50Hz_nonlinear}(b)) we clearly recognize two additional spectral signatures at the second harmonic. The one overlapping the longitudinal mode corresponds to the homogeneous component, while the other which lies precisely on the dashed curve, is unequivocally identified as the forced component. To the authors' best knowledge, this is the first experimental study that explicitly reveals the co-existence of the two distinct nonlinear contributions in a two-dimensional periodic structure.

     From inspection of the band diagram, we have already observed that the fundamental and second harmonics feature distinct and complementary modal characteristics. The response at the fundamental harmonic, while blending shear and longitudinal effects, is dominated by the shear, which represents a softer mechanism. In contrast, the response at the second harmonic (more precisely, the homogeneous part, for which a clear modal structure can be invoked) is dominated by longitudinal mechanisms. By this observation, we can characterize this result as an instance of modal enhancement, whereby the nonlinear activation of the second harmonic introduces in the response modal characteristics that are complementary to those exhibited by the linear response. The vibrometer scan also allows exploring the manifestation of SHG on the spatial pattern of the wave response. In Fig.~\ref{Wavefield} we plot six snapshots of the propagating wavefield. In Fig.~\ref{Wavefield}(a), we show the total wavefield, encompassing fundamental and second harmonics, while in Fig.~\ref{Wavefield}(b) and (c) we isolate the filtered components at $\nu_0$ and $2\nu_0$, respectively. The wavefields in Fig.~\ref{Wavefield} reveal spatial complementarity between the predominantly vertical directivity of the second harmonic and the quasi-isotropic pattern of the fundamental harmonic. This result indicates that, in addition to modal enrichment, nonlinearity is also responsible for a directivity enrichment. 
        
    \begin{figure} [!htb]
    	\centering
    	\includegraphics[scale=0.27]{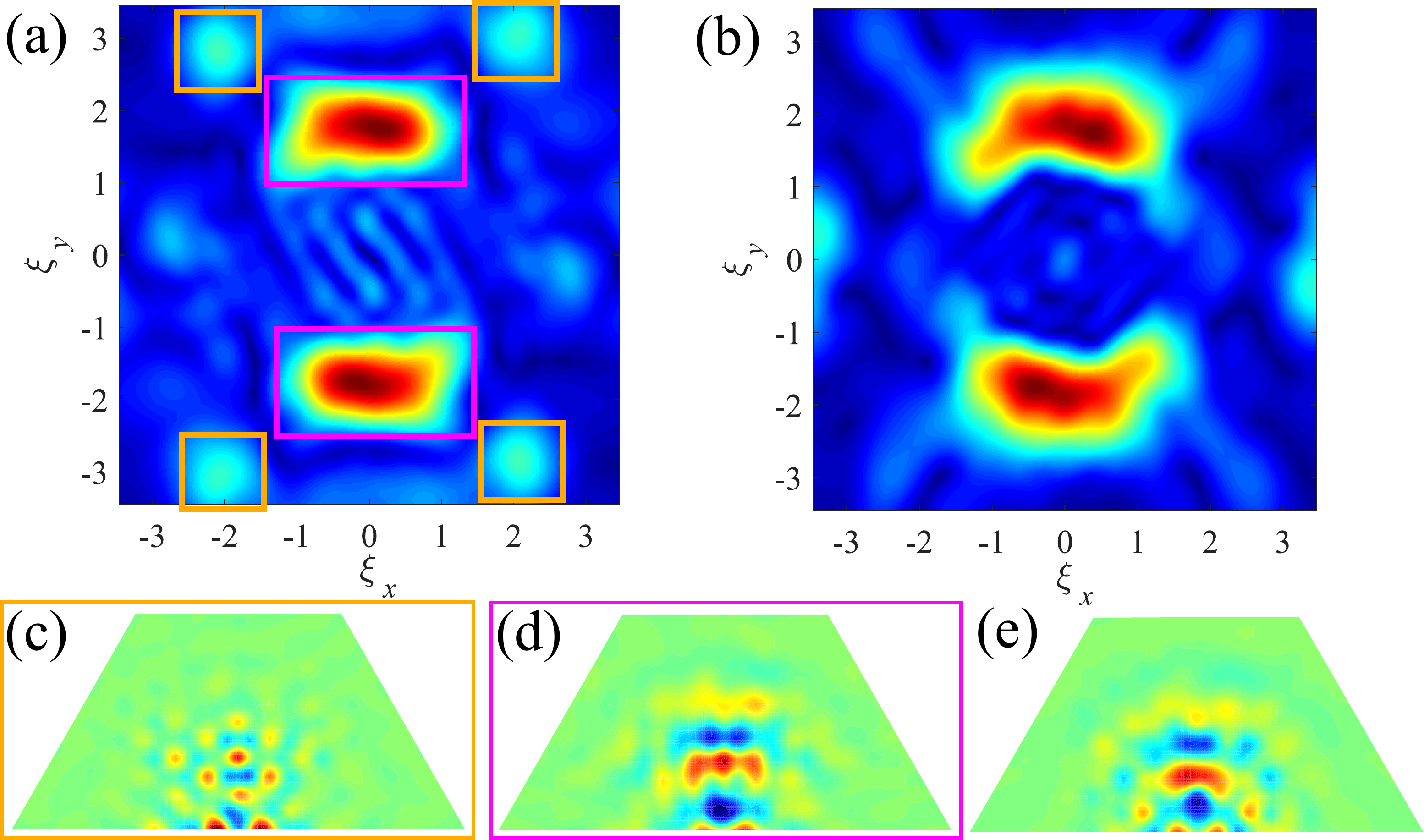}
    	\caption{$\mathbf{k}$-plane amplitude spectra of (a) the filtered second harmonic of a nonlinear response and (b) a linear wavefield excited directly at $2\nu_0$. (c) and (d) Filtered wavefields of the spectral components inscribed in amber and magenta boxes, respectively.}
    	\label{Spectrum_2DXY}
    \end{figure}

    To highlight separately and filter the spatial contributions of the homogeneous and forced components that coexist at the second harmonic, we subject the last snapshot of the wavefield filtered at $2\nu_0$ (i.e., the last snapshot in Fig.~\ref{Wavefield}(c)) to 2D-DFT in space. The resulting $\mathbf{k}$-space spectrum is plotted in Fig.~\ref{Spectrum_2DXY}(a). For comparison, we repeat the exercise for a linear wavefield obtained with a low-amplitude excitation prescribed directly at $2\nu_0$ (Fig.~\ref{Spectrum_2DXY}(e)), whose spectrum is shown in Fig.~\ref{Spectrum_2DXY}(b). In Fig.~\ref{Spectrum_2DXY}(a), we identify two spectral signatures at the second harmonic (appearing with their mirror counterparts due to the symmetry folding operations involved in the 2D-DFT). The dominant component, inscribed by magenta boxes, is consistent with the spectrum of the linear wavefield excited directly at $2\nu_0$ (i.e., Fig.~\ref{Spectrum_2DXY}(b)), and is therefore interpreted as the homogeneous component. On the other hand, the secondary contribution, inscribed by amber boxes, which is germane to the nonlinear response, must be interpreted as the forced response. By zeroing out the two signatures, one at a time, and carrying out an inverse 2D-DFT of the remainder, we can filter out the separate wavefields of the two components, as shown in Fig.~\ref{Spectrum_2DXY}(c) and (d), respectively. Clearly, the pattern in the magenta box of Fig.~\ref{Spectrum_2DXY}(d), which exhibits longitudinal behavior, is reminiscent of the wavefield in Fig.~\ref{Spectrum_2DXY}(e), further supporting, from a spatial perspective, the notion that the homogeneous component conforms to the linear response that would be observed in the lattice if the excitation were prescribed directly at $2\nu_0$. The other, shown in Fig.~\ref{Spectrum_2DXY}(c) and corresponding to the forced component, displays a more dispersive spatial pattern. It is worth emphasizing again that, while it is possible that the homogeneous component may be in part triggered by nonlinearities extrinsic to the mechanical system and due, for instance, to harmonics buried in the excitation signal, the forced component is germane to the SHG established within the structure. Therefore, the existence of the force second harmonic is the most powerful detector of nonlinearities in the lattice. 

    In summary, we have experimentally characterized the nonlinear wave response of a lattice of repulsive magnets on an elastic foundation. First, we have demonstrated the existence of SHG in the specimen, separately pinpointing its two second harmonic contributions. Then, we have shown that the nonlinear response features modal characteristics that are complementary to the those of the fundamental wave. Finally, we have reconstructed the spatial characteristics of the two nonlinear components, revealing additional complementarity between them in terms of spatial characteristic and directivity. The magnetic lattice prototype reveals to be an ideal platform for the experimental observation of nonlinear wave propagation. On one hand, its compliance is conducive to displacements that are at least one order of magnitude larger than what is achievable in hard solid specimens and approaching the advantages of soft materials without their pitfalls in terms of damping. On the other hand, its discrete nature results in low modal complexity, compared to a structural lattice, thus facilitating a clear identification of all the spectral components. 
	
	\section*{Acknowledgement}
	This work is supported by the National Science Foundation (CAREER Award CMMI-$1452488$). The authors are indebted to Lijuan Yu for her precious help with the specimen assembly and to Joseph Labuz, Xiaoran Wang and Chen Hu for sharing their invaluable expertise with the force-displacement testing apparatus.  
	
	\bibliography{myrefs}
	
\end{document}